\documentclass[usenatbib]{mn2e}
\usepackage{graphicx}

\newcommand\E[1]{\times10^{#1}}
\newcommand\un[1]{{\,\rm #1}}
\newcommand\rs[1]{_\mathrm{#1}}
\newcommand\g{$\gamma$}
\newcommand\ApJ{ApJ}

\title[Hadronic $\gamma$-ray images of SNRs. SN~1006.]{Hadronic $\gamma$-ray images of Sedov supernova remnants}
\author[Beshley V. et al.]{V.~Beshley$^{1}$, O.~Petruk$^{1,2}$\\
$^{1}$Institute for Applied Problems in Mechanics and Mathematics, Naukova St.\ 3-b,
   79060 Lviv, Ukraine\\
$^{2}$Astronomical Observatory, National University, Kyryla and Methodia St.\ 8, 79008 Lviv, Ukraine\\
}

\begin{document}

\date{Accepted .... Received ...; in original form ...}

\pagerange{\pageref{firstpage}--\pageref{lastpage}} \pubyear{2011}

\maketitle

\label{firstpage}

\begin{abstract}
A number of modern experiments in high-energy astrophysics produce images of supernova remnants (SNRs) in the TeV and GeV gamma-rays. Either relativistic electrons (due to the inverse-Compton scattering) or protons (due to the pion decays) may be responsible for this emission. In particular, the broad-band spectra of SNRs may be explained in both leptonic and hadronic scenarios. Another kind of observational data, namely, images of SNRs, is an important part of experimental information. We present a method to model gamma-ray images of Sedov SNRs in uniform media and magnetic field due to hadronic emission. 
These \g-rays are assumed to appear as a consequence of meson decays produced in inelastic collisions of 
accelerated protons with thermal protons downstream of the shock -- a model would be relevant for SNRs without firm confirmations of the shock-cloud interaction, as e.g. SN~1006. 
Distribution of surface brightness of the shell-like SNR is synthesized numerically for a number of configurations. An approximate analytical formula for azimuthal and radial variation of hadronic \g-ray brightness close to the shock is derived. 
The properties of images as well as the main factors determining the surface brightness distribution are determined. 
Some conclusions which would be relevant to SN~1006 are discussed.
\end{abstract}


\section{Introduction}

Cosmic rays (CRs) are an important component of the Universe on different scales, from the Solar System to clusters of galaxies. 
Supernova remnants (SNRs), sources of galactic CRs, are excellent objects to study magneto-hydrodynamics of nonrelativistic shocks and acceleration of cosmic rays, as well as their mutual influence. Protons and electrons, once heated and accelerated on the shock, produce different types of emission: thermal with maximum intensity in X-rays and nonthermal emission in radio, X-rays and \g-rays. Experiments in high-energy astronomy observe all types of these emission. 

Most of galactic cosmic rays are believed to be produced by the forward shocks in supernova remnants (SNRs). A number of indirect evidences favor this expectation. In particular, efficient proton acceleration changes the structure of the shock front and makes plasma more compressible that leads to lower adiabatic index, to increased shock compression ratio and to some observed effects: reduced physical separation between the forward shock and the ``contact discontinuity'' (or reverse shock) \citep[e.g.][]{warren-et-al-2006} or even protrusions of the ejecta clumps beyond the forward shock \citep{rakowski-et-al-2011}; concave shape of the energy spectrum \citep[e.g.][]{reynolds-ellison-1992}; growth of some turbulence modes and amplification of magnetic field in the pre-shock region \citep[e.g.][]{Bell2004}; ``blinking'' X-ray spots originated from such growth of magnetic field  \citep{Uchiyama2007,patnaude07}; ordered non-thermal X-ray strips \citep{eriksen2011,bykov2011}. 

Nevertheless, there is still luck of direct observational confirmations that protons are accelerated in SNRs to energies which make them responsible for \g-ray emission of SNRs. Broadband spectra of a number of SNRs may be explained either by leptonic or hadronic scenario for \g-ray emission \citep[e.g. SN~1006:][]{HESS-SN1006-2010}. Fermi \g-ray observatory is expected to clarify such ambiguity, at least in  bright SNRs \citep[as in RX~J1713.7-3946:][]{RX1713abdo_et_al-2011}. 
In any case, other kind of experimental information should be studied as well. Namely, the distribution of surface brightness in different bands caused by emission of accelerated particles in SNRs contains wealth of information about properties of CRs and magnetic fields in these objects. 

A well known example is the method to estimate the strength of the post-shock magnetic field from the thickness of the radial profiles of X-ray brightness \citep[e.g.][]{Ber-Volk-2003-mf}. Radial structure of the shock front upstream (observed in X-rays) seems to confirm back-reaction of particles and magnetic field amplification, locally, in SN~1006 \citep{Morlino-etal-2010}. A method to derive an aspect angle (between ambient magnetic field and the line of sight) in SNR from the azimuthal variation of the radio brightness is developed in \citet{pet-SN1006mf} and applied to SN~1006 under assumption of the uniform interstellar magnetic field (ISMF). More detailed consideration of the azimuthal radio profile including the ISMF nonuniformity results in constraints on orientation of ISMF and its gradient around SN~1006 \citep{FB-et-al-2011}. Radial profiles of the radio brightness may constrain the time evolution of the electron injection efficiency \citep{SN1006cp}. Spatially resolved spectral analysis of radio and X-ray data \citep[as in SN~1006:][]{SN1006Marco} determines the model, value and surface variation of the electron maximum energy in the SNR \citep{SN1006cp}. 

Properties of the nonthermal images of Sedov SNRs due to radiation of accelerated electrons in radio, X-rays and \g-rays are systematically studied in \citet{Reyn-98,Reyn-04} and \citet[][Papers I and II respectively]{thetak,xmaps}. Numerical models for synthesis of maps of adiabatic SNRs in uniform interstellar medium (ISM) and uniform ISMF from basic theoretical principles as well as approximate analytical descriptions are developed in these papers. The main factors determining the azimuthal and radial variation of surface brightness of SNRs are determined. The role of the nonuniform ISM and/or nonuniform ISMF in nonthermal images of SNRs are studied by \citet{Orletal07,Orletal11}; in particular, gradients of ISMF strength or ISM density result in several types of asymmetries in surface brightness distributions of SNRs in radio, hard X-ray and \g-ray bands. 
These papers are limited to the test-particle approach because the non-linear theory of diffusive acceleration is not developed for shocks of different obliquity, while the obliquity dependence of various parameters is crucial in image modeling. 

In addition to the analysis of SNR images simulated from basic theoretical principles, model-independent methods for SNR images are important.
Such method for synthesis of the inverse-Compton \g-ray map of SNR from the radio (or hard X-ray) image and results of the spatially resolved X-ray spectral analysis is developed and applied to SN~1006 in \citet{Petetal09icp}. 
It is found that synthesized inverse-Compton \g-ray image of SN~1006 
is in agreement with HESS observations. 
This fact favors a leptonic scenario for the TeV 
\g-ray emission of this SNR. Further development of this method allows us to find a new way to constrain the strength of magnetic field in SN~1006 from its nonthermal images \citep{pet-kuzyo-fb-2011}. 

Though a leptonic scenario for \g-rays from SN~1006 is reasonable and promising \citep[e.g.][]{SN1006cp}, a hadronic origin 
cannot be ruled out even in view of the small ISM 
densities \citep[upper limit is $0.5\un{cm^{-3}}$; e.g.][]{Dubner_et_al_2002,acero_et_al2007}, 
which are consistent with a hadronic scenario \citep{Ber-Volk-2009-sn1006,HESS-SN1006-2010}. 
The shape of the observed spectrum in the TeV \g-ray range better corresponds 
to the hadronic spectrum rather then to leptonic one \citep{HESS-SN1006-2010}. 
In order to fit the observed radio, X-ray and gamma-ray emission within the leptonic scenario
one needs rather high downstream magnetic field $\approx 30\un{\mu G}$ \citep{Volk-Ber-2008,SN1006cp}; 
there is a general thought that such a field can only 
be produced by efficiently accelerated cosmic ray proton component.

If TeV \g-rays from SN~1006 is of hadronic origin, 
then the observed brightness map 
of this SNR in \g-rays with energy $>1\un{TeV}$ 
should reflects the distribution of protons with energies
$> 2\un{TeV}$ which interact with shocked thermal protons inside SNR. 

In the present paper, we study this possibility modeling \g-ray images of a Sedov SNR in uniform ISM and uniform ISMF it would have under such scenario. 
Though we are primarily interested in comparison with SN~1006, our results are general and may be used for analysis of other SNRs.

\section{Model}
\subsection{General description}

Our model closely restores model used in Papers I and II. 
Let us consider an adiabatic SNR in uniform ISM and uniform ISMF. 
Hydrodynamics of the remnant is given by the self-similar \citet{Sedov-59} solutions; in practice, we use their quite accurate approximations in Lagrangian coordinates \citep[Sect.~4 in][]{petr2000}. 
Magnetic field is described following \citet{Reyn-98}; 
we do not consider amplification of the ambient field (though its role is discussed in Sect.~\ref{pmaps:approx_form}). 
Accelerated protons are described by a number of parameterizations as follows. 

At the shock, accelerated protons are distributed with energy $E\rs{p}$ as  
\begin{equation}
 N(E\rs{p}) =K\rs{s}E\rs{p}^{-s+\delta s(E\rs{p})}\exp\left(-\left(E\rs{p}/E\rs{p,max}\right)^{\alpha}\right), 
 \label{pmaps:NE}
\end{equation}
where $N$ is the differential number density of accelerated protons, 
$E\rs{p,max}$ is the maximum energy of protons, $\delta s(E\rs{p})$ reflects possible concavity of the spectrum shape as an effect of efficient acceleration, $s$ and $\alpha$ are constants, $K\rs{s}$ the normalization, index `s' denotes values at the shock. 
Different theoretical approaches results in different values of $\alpha$ \citep[see references in][]{Reynolds-Keoh-99,allen2008,Orletal11,xmaps} which are 
between 1/4 and 2. 
The function $\delta s(E\rs{p})$ varies a bit and very slowly over decades in energy $E\rs{p}$ \citep[e.g.][]{Ber-Ell-simple_model}. The emission at some photon energy $\varepsilon$ is mostly determined by the narrow range of $E\rs{p}$. Therefore, $\delta s(E\rs{p})$ 
produces negligible changes to our results and we take it $\delta s(E\rs{p})=0$. 

The maximum energy may vary with obliquity: 
\begin{equation}
 E\rs{p,max}(t,\Theta\rs{o})=E\rs{p,max\|}(t) {\cal E}\rs{max}(\Theta\rs{o}). 
\end{equation} 
where 
$E\rs{p,max\|}(t)\propto V(t)^{q}$, $V$ the shock velocity, $q$ a constant, 
index ``$\|$'' denotes values at parallel shock, 
$\Theta\rs{o}$ is the obliquity angle between the ambient magnetic field and the shock velocity, 
function ${\cal E}\rs{max}(\Theta\rs{o})$ is some function. 

There are expectations that the efficiency of injection (defined as the ratio of density of accelerated protons to density of all protons) 
could be most efficient at the parallel shock and progressively depressed toward regions of SNR surface where the shock is perpendicular 
\citep{ell-bar-jones-95,Volk-Ber-2003-inj}. In opposite, an analysis of known SNR images presents some hints against such scenario \citep{reyn-fulbr-90,Orletal07}; it also reveals that properties of SNR images are similar in isotropic (no dependence on obliquity) and quasi-perpendicular (injection prefers perpendicular shock) approaches. In our calculations, we use simple parameterizations for obliquity dependence of the injection. 
Namely, the obliquity dependence of the normalization $K\rs{s}$ (which is proportional to the injection efficiency) is parameterized by 
\begin{equation}
 K\rs{s}(t,\Theta\rs{o}) = K\rs{s\parallel}(t){\cal K}(\Theta\rs{o})
\end{equation}
where ${\cal K}(\Theta\rs{o})$ is given by 
the following formulae where different `sensitivity' to the obliquity angle is given by the parameter $\Theta\rs{K}$. 
The quasi-parallel and quasi-perpendicular injections are represented respectively by 
\begin{equation}
 {\cal K}(\Theta\rs{o})=
 \exp\left(-\big({\Theta\rs{o}}/{\Theta\rs{K}}\big)^2\right),
 \label{finj}
\end{equation}
\begin{equation}
 {\cal K}(\Theta\rs{o})=
 \exp\left(-\big({(\Theta\rs{o}-\pi/2)}/{\Theta\rs{K}}\big)^2\right).
 \label{finjperp}
\end{equation}
Isotropic injection ${\cal K}(\Theta\rs{o})=\mathrm{const}$ may be restored by a large value of $\Theta\rs{K}$. 
Evolution of the injection efficiency may be accounted through $K\rs{s\|}\propto V(t)^{-b}$ where $b$ is a parameter.

The evolution of the proton energy spectrum in the SNR's interior is considered in the next subsections. 

The surface brightness is calculated integrating emissivities along the line 
of sight within SNR. 
Hadronic \g-rays appear as a consequence of the neutral pion and $\eta$-meson decays produced in inelastic collisions of 
accelerated protons with {\em thermal protons downstream of the shock}; the spatial distribution of the target protons inside the volume of SNR is simply 
proportional to the local plasma density. 
We use the approach of \citet{Aha-Ato-2000} to calculate the hadronic $\gamma$-ray emissivity: 
\begin{equation}
 q_{\gamma}(\varepsilon)=2\int\frac{q_{\pi}(E_{\pi})dE_{\pi}}{\sqrt{E_{\pi}^{2}-m_{\pi}^{2}c^{4}}},
 \label{pp-emiss}
\end{equation}
where $E_{\pi}$ and $m_{\pi}$ are the energy and mass of the neutral pion, the pion emissivity is
\begin{equation}
 q_{\pi}(E_{\pi})=\frac{cn\rs{H}}{\kappa}
 \sigma\rs{pp}\left(E\rs{p}'\right)N\left(E\rs{p}'\right),
\end{equation}
the average energy of protons mostly responsible for creation of pions with energy $E\rs{\pi}$ is
\begin{equation}
 E\rs{p}'=m\rs{p}c^2+{E_{\pi}}/{\kappa},
\end{equation}
$\kappa=0.17$ (this value accounts also for $\eta$-meson production), the cross-section of inelastic collision of a highly-energetic proton with a thermal proton (with negligible energy comparing to the energy of the incident proton) is 
\begin{equation}
 \sigma\rs{pp}(E\rs{p})=28.5+1.8\ln{(E\rs{p}/1\un{GeV})} \ \  \un{mb}.
 \label{across}
\end{equation}


\subsection{Proton energy losses due to meson production}

In order to synthesize SNR images, we need to know emissivity (\ref{pp-emiss}) in each point of the interior and, therefore, to describe evolution of the proton energy spectrum. Relativistic protons lose their energy due to adiabatic expansion and inelastic collisions. 
Energy losses of protons in the meson production is (Appendix \ref{pmaps:app0})
\begin{equation}
 -\left(\frac{dE\rs{p}}{dt}\right)\rs{pp}=3\kappa cn\rs{H}\sigma \rs{pp}(E\rs{p})E\rs{p,kin},
 \label{Bloss3_3}
\end{equation}
where $c$ is the speed of light, $n\rs{H}$ the number density of the target protons, the factor $3$ accounts for the production of $\pi^{0}$, $\pi^{+}$ and $\pi^{-}$ mesons. 

The losses due to proton collisions are more significant for larger density of the target protons and higher energy of incident protons. In order to feel then the proton energy losses are effective, let us compare them with the radiative energy losses of electrons 
\begin{equation}
 -\left(\frac{dE\rs{e}}{dt}\right)\rs{rad}=\frac{4}{3}\sigma\rs{T}c\left(\frac{E\rs{e}}{m\rs{e}c^{2}}\right)^{2}\left(\frac{B^{2}}{8\pi}\right)
\end{equation}
where $\sigma\rs{T}$ is the Thomson cross-section, $m\rs{e}$ the mass of electron. 
The ratio of electron to proton losses is
\begin{equation}
 \frac{\dot{E}\rs{e,rad}}{\dot{E}\rs{p,pp}}\simeq 5.0\ \frac{B^{2}\rs{\mu G}\ E^{2}\rs{e,TeV}}{n\rs{H}\ E\rs{p,TeV}},
 \label{ploss}
\end{equation}
where we used $\sigma\rs{pp}\approx 33\un{mb}$ \citep{Aha-Ato-2000}, $B\rs{\mu G}$ magnetic field in $10^{-6}\un{G}$, $E\rs{e,TeV}$ and $E\rs{p,TeV}$ are the energies of electrons and protons measured in $10^{12}\un{eV}$. One can see that 
the losses of protons with energy $100$ TeV are comparable to losses of electrons with energy $30$ TeV in magnetic field $30\un{\mu G}$, if 
the number density of target protons is $4\E{4}\un{cm^{-3}}$ respectively. 

\subsection{Downstream evolution of the proton energy distribution}

Let the energy of proton at the time $t\rs{i}$, when it leaves the region of acceleration, was $E\rs{pi}$. Then it is smaller at the present time $t$ (Appendix \ref{pmaps:app1}),
\begin{equation}
 E\rs{p}={E\rs{pi}}{{\cal E}\rs{ad}(\bar{a})^{\mu(\bar{a})}{\cal E}\rs{pp}(E\rs{p},\bar{a})}, 
\end{equation}
because the terms responsible for the adiabatic ${\cal E}\rs{ad}$ and collisional ${\cal E}\rs{pp}$ losses are equal or smaller than unity; $\bar a=a/R$, $a$ the Lagrangian coordinate, $R$ the radius of SNR,
\begin{equation}
 {\cal E}\rs{ad}(\bar{a})=\bar n(\bar a)^{1/3},
 \label{ppmaps:adloss}
\end{equation}
where 
$\bar n=n/n\rs{s}$, 
index ``s'' denotes the value immediately post-shock, 
\begin{equation}
 {\cal E}\rs{pp}(E\rs{p},\bar{a})=\left(E\rs{p}/1\un{GeV}\right)^{1-\mu(\bar{a})}{\cal I}(\bar{a}),
 \label{adpploss_t}
\end{equation}
$\mu(\bar{a})$ and ${\cal I}(\bar{a})$ are dimensionless self-similar functions presented in the Appendix \ref{pmaps:app1}. 
Close to the shock they behave as (Appendix \ref{pmaps:app2})
\begin{equation}
 {\cal E}\rs{pp}(\bar{a})\approx\bar a^{\zeta\sigma\rs{pp}(E\rs{p})},
 \qquad 
 \mu(\bar a)\approx \bar a^{-1.8\zeta}
 \label{ppmaps:adloss_app}
\end{equation}
where the cross section is in mb, $\zeta=1.21\E{-6}t_3n\rs{Hs}$, $t_3=t/1000\un{yrs}$, $t$ the age of SNR. One can note that ${\cal E}\rs{pp}$ is effective if the number density of the target protons or/and age of SNR are large. 

The energy spectrum of protons downstream of the Sedov shock evolves in a self-similar way (Appendix \ref{pmaps:app1})
\begin{equation}
\begin{array}{l}
 \displaystyle
N\rs{p}(E\rs{p},\bar a,t)=K(\bar a,t)E^{-s}\rs{p}\ \mu(\bar a){\cal E}\rs{pp}(E\rs{p},\bar{a})^{s-1}
\\ \\ \displaystyle\qquad
\times\exp\left[ -\left(\frac{E\rs{p}\bar{a}^{3q/2}}{E\rs{p,max}{\cal E}\rs{ad}(\bar{a})^{\mu(\bar{a})}{\cal E}\rs{pp}(E\rs{p},\bar{a})} \right)^{\alpha}\right] .
\end{array} 
 \label{specev_}
\end{equation}
with $K(\bar a,t)=K\rs{s}\bar{K}(\bar{a})$, $\bar{K}(\bar{a})=\bar{a}^{3b/2}\bar{n}(\bar{a})^{1+\mu(\bar{a})(s-1)/3}$.
The downstream distribution of relativistic protons are modified mainly by the adiabatic expansion of SNR. 
Losses due to inelastic collisions affects the distribution only when $\zeta$ is not small, that happens when $n\rs{Hs}$ and/or $t$ are large. 

\begin{figure*}
 \centering
 \includegraphics[width=16truecm]{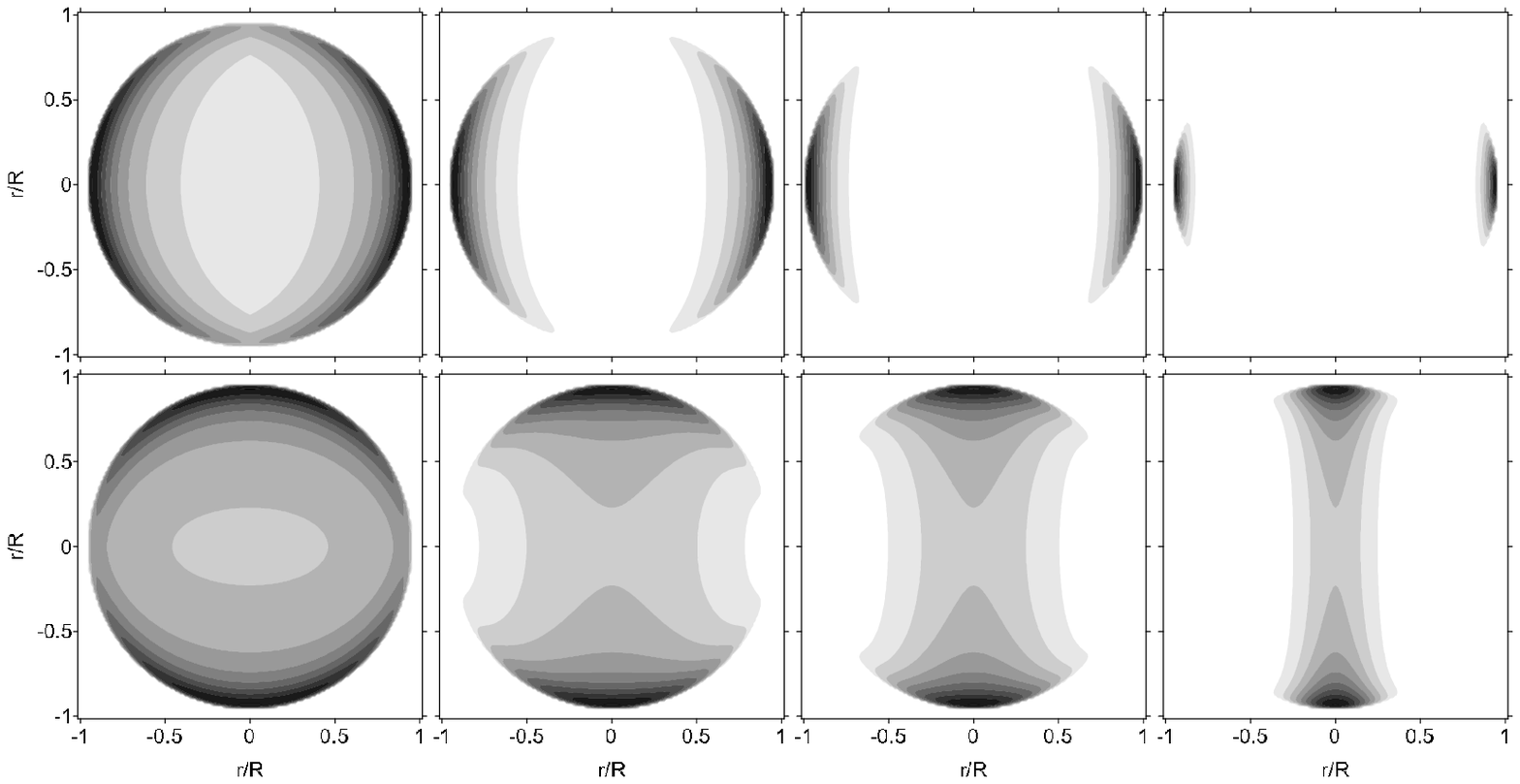}
 \caption{Hadronic $\gamma$-ray images of Sedov SNR for different models of injection: quasi-parallel (upper panel) 
 and quasi-perpendicular (lower panel) for an aspect angle $\phi\rs{o}=90^\mathrm{o}$. 
 Parameter $\Theta_{K}$ is $\pi/2$, $\pi/4$, $\pi/6$, $\pi/12$ (from left).
 Hereafter, the ISMF is parallel to the horizontal axis and the increment in brightness is $\Delta S=0.1 S\rs{\max}$.
 }
 \label{fig1}
\end{figure*}
\begin{figure*}
 \centering
 \includegraphics[width=16truecm]{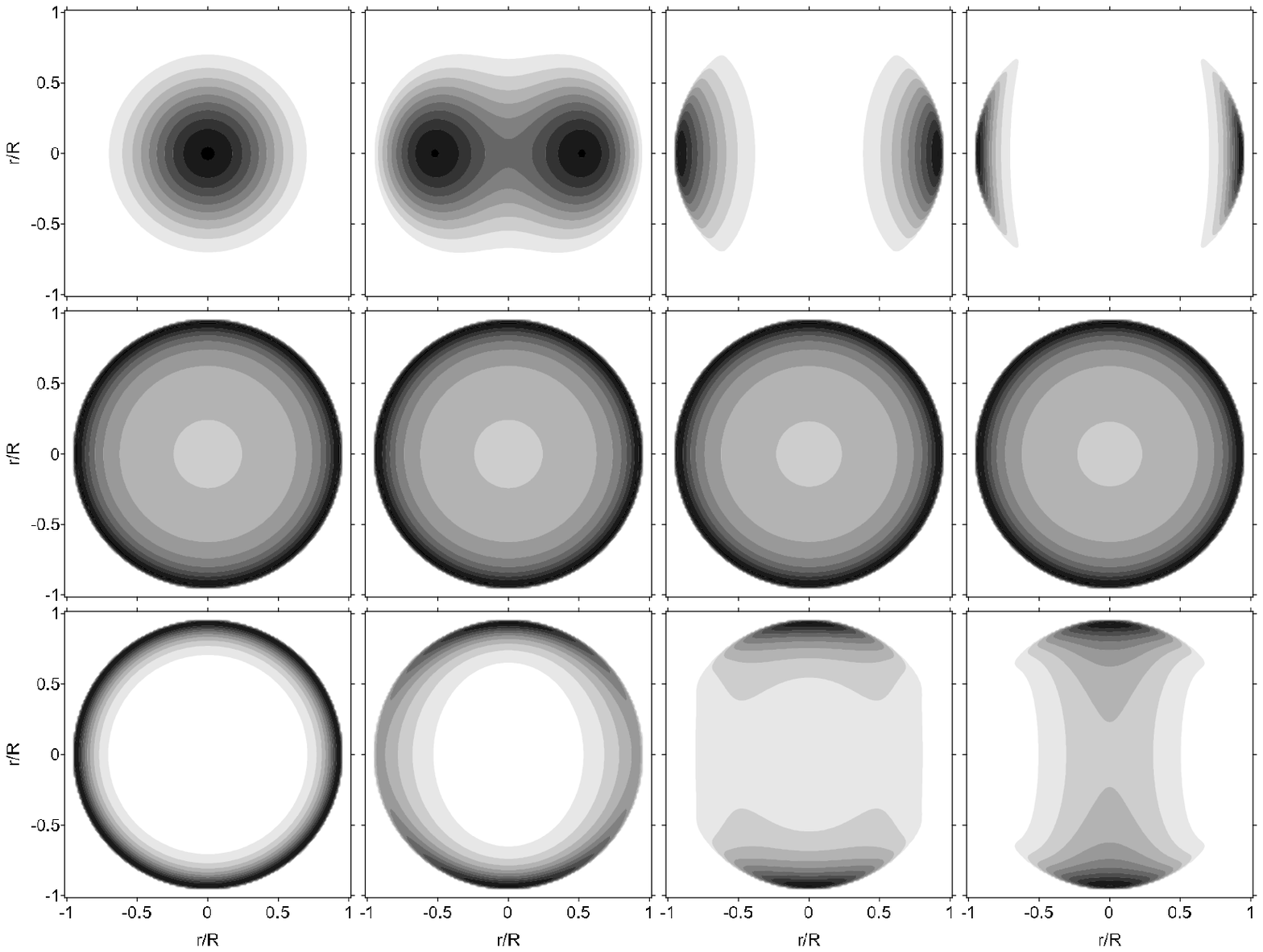}
 \caption{The same as on Fig.~\ref{fig1} for different aspect angles. 
 Models of injection are quasi-parallel (top, $\Theta_{K}=\pi/6$), isotropic (centre) and quasi-perpendicular (bottom, $\Theta_{K}=\pi/6$).
 Aspect angle $\phi\rs{o}=0^\mathrm{o}$, $30^\mathrm{o}$, $60^\mathrm{o}$ and $90^\mathrm{o}$ (from left). 
 }
 \label{fig2}
\end{figure*}

\section{General properties of hadronic images. Numerical simulations}
\label{sect-numeric}

In order to narrow the parameter space, the images of SNRs in the present section are synthesized using the following set of parameters. Namely, we take $\gamma=5/3$, $\delta s=0$, $s=2$, $\alpha=1$,  $K$ constant in time (i.e. $b=0$), the energy of \g-rays $\varepsilon=1\un{TeV}$. 
The actual representation for ${\cal E}\rs{max}(\Theta\rs{o})$ used in our calculations is monotonic function which corresponds to the 
time-limited model of \citet{Reyn-98} with so called ``gyrofactor'' $\eta=1$ (Fig.~\ref{fig1}, \ref{fig2}, \ref{fig5}) or $\eta=5$ (Fig.~\ref{fig3}, \ref{fig6}); 
such form provides $q\approx 0$ and $E\rs{p,max}$ at the perpendicular shock to be 2 or 26 times the value at the parallel shock. 
The age of SNR and the number density are taken $t=1000\un{yrs}$ and $n\rs{Hs}=1\un{cm^{-3}}$ except of Fig.~\ref{fig5} where the densities are given in the caption\footnote{parameter really influencing a \g-ray image of SNR is $\zeta$, a product of $t$ and $n\rs{Hs}$, Eq.~(\ref{ppmaps:adloss_app}).}. 
Role of other choices of the above parameters is visible from the approximate formula presented in Sect.~\ref{pmaps:approx_form}. 

Radio and X-ray synchrotron images of SNRs depends directly on the distribution of magnetic field. Magnetic field affects the TeV \g-ray images of SNR (due to inverse-Compton emission) indirectly, through modification of the downstream evolution of emitting relativistic electrons, namely, the larger the field the larger the radiative losses and, therefore, the thinner the region behind the shock occupied by the electrons able to emit \g-rays in the TeV band (Papers I and II). 
Magnetic field appears in the hadronic \g-rays only through dependence (if any) of the injection efficiency and the maximum energy on the shock obliquity. Therefore, the azimuthal variation of brightness in such images depends mainly on the two functions: ${\cal K}(\Theta\rs{o})$ and ${\cal E}\rs{max}(\Theta\rs{o})$. 

\subsection{Different models of injection}

Fig.~\ref{fig1} shows how $\Theta_{K}$ affects $\gamma$-ray image of SNR it has due to interactions of accelerated protons with thermal protons downstream of the shock. Different dependence of the injection efficiency on obliquity results in two different types of SNR morphology (Fig.~\ref{fig1}). Quasi-parallel injection  creates the 'polar-caps' structure (the lines of ambient magnetic field cross the bright limbs) while the isotropic and the quasi-perpendicular injections are responsible for the 'barrel-shaped' remnant (ambient magnetic field is parallel to limbs in this case). Increase of $\Theta_{K}$ results in decrease of the azimuthal width of the limbs due to the progressive luck of accelerated protons. 

The role of the aspect angle $\phi\rs{o}$ (between the ambient magnetic field and the line of sight) is shown on Fig.~\ref{fig2}. Spherically-symmetric morphology transforms to bilateral one with increasing the angle if the proton injection depends on obliquity.

\subsection{Dependence on the maximum energy}

\begin{figure}
 \centering
 \includegraphics[width=8.0truecm]{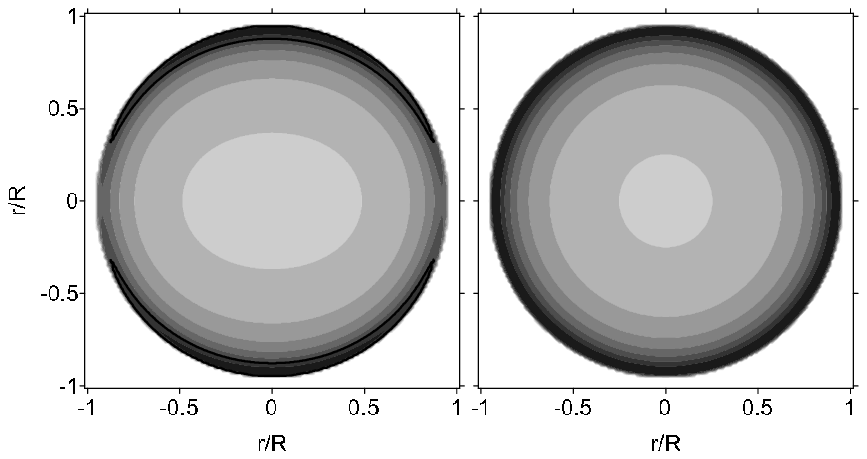}
 \caption{Role of $E\rs{max}$ in hadronic \g-ray images of Sedov SNR. 
 $E\rs{max\|}=10\un{TeV}$ (left, $\epsilon\rs{m}=0.82$), $E\rs{max\|}=1000\un{TeV}$ (right, $\epsilon\rs{m}=0.0082$). 
 Isotropic injection, ${\cal E}\rs{max}(\pi/2)/{\cal E}\rs{max}(0)=26$, $\alpha=1$, 
 aspect angle $\phi\rs{o}=90^\mathrm{o}$.
  }
 \label{fig3}
\end{figure}

The ratio of $E\rs{max}$ at the perpendicular and parallel shock is 2 on previous figures. 
Let the maximum energy of protons increase more rapidly from parallel to perpendicular shock, in particular in 26 times as on Fig.~\ref{fig3}. 
In this case, the azimuthal variation of $E\rs{max}$ is in general more prominent. 

The thickness of limbs depends on two factors.  First, on the ratio $\epsilon\rs{m}=E\rs{pm}/E\rs{p,max\|}$ where $E\rs{pm}$ is the energy of proton which effectively gives the most contribution to \g-rays with energy $\varepsilon$; it is given by Eq.~(\ref{emin}); once we are interested in $\varepsilon> 1\un{GeV}$, it may be simplified to 
\begin{equation}
 E\rs{pm}(\varepsilon)=\varepsilon{\xi}/{\kappa}=8.2\varepsilon.
\end{equation}
where we used $\xi=1.4$ (Appendix \ref{pmaps:app3}). Namely, if 
$\epsilon\rs{m}\ll 1$ then the obliquity variation of the maximum energy does not affect the image of SNR (Fig.~\ref{fig3}). 
Following this conclusion, we would like to note that the role of $E\rs{max}$ may be prominent in the analysis of the Fermi observations ($\varepsilon\sim 3\un{GeV}$) only if $E\rs{p,max}$ is smaller than $\sim 25\un{GeV}$ that is unreasonable. Therefore, the role of $E\rs{p,max}$ may be neglected during analysis of observations in GeV \g-rays.

\begin{figure}
 \centering
 \includegraphics[width=8.0truecm]{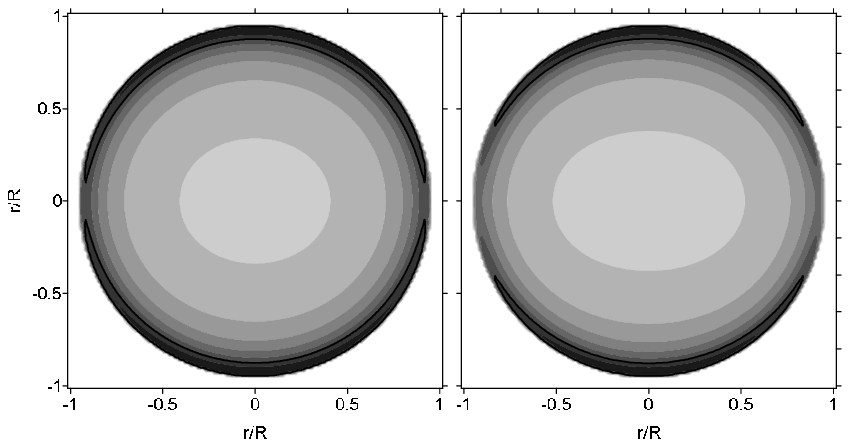}
 \caption{The same as on Fig.~\ref{fig3} for $\alpha=0.5$ (left) and 2 (right); 
 $E\rs{max\|}=10\un{TeV}$.}
 \label{fig6}
\end{figure}

\begin{figure}
 \centering
 \includegraphics[width=8.0truecm]{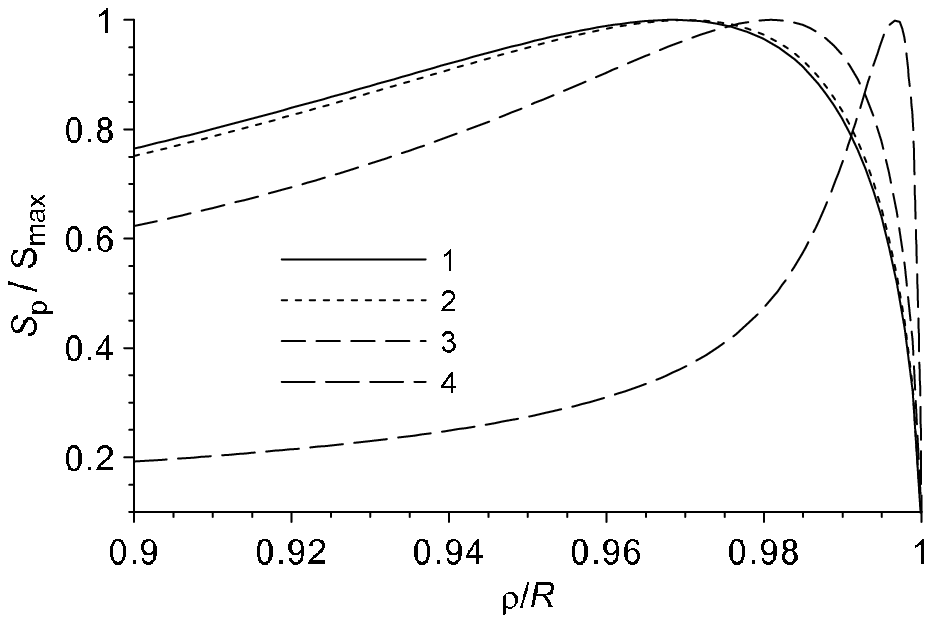}
 \caption{Radial profiles of \g-ray surface brightness due to hadronic emission for different densities $n\rs{Hs}$ of the target protons: 
 $1\un{cm^{-3}}$ (line 1), 
 $10^{4}\un{cm^{-3}}$ (line 2),
 $10^{5}\un{cm^{-3}}$ (line 3),
 $10^{6}\un{cm^{-3}}$ (line 4). 
 Isotropic injection, ${\cal E}\rs{max}(\Theta\rs{o})=\mathrm{const}$, $E\rs{max\|}=1000$ TeV, $\phi\rs{o}=90^\mathrm{o}$, 
 azimuth angle (measured from direction of the ambient magnetic field) $\varphi=0^\mathrm{o}$.
 $S\rs{max}$ is the peak value of $S\rs{p}(\rho)$.
 }
 \label{fig5}
\end{figure}

The second factor which affect the azimuthal thickness of limbs is how quickly the proton energy spectrum $N(E\rs{p})$ ends (given by the parameter $\alpha$ in our model). Fig.~\ref{fig6} demonstrates this effect: the larger $\alpha$ the more rapid decrease of the high-energy end of the proton spectrum and the smaller the azimuthal extension of the limbs. 

\subsection{Role of density of the target protons}

Energy losses of relativistic protons due to collisions with thermal protons are proportional to the density of the target protons, Eq.~(\ref{Bloss3_3}).
They were negligible in the calculations presented above because the number density was $n\rs{Hs}=1\un{cm^{-3}}$. 
The collisional losses of protons are important for much higher densities as it is evident from Eq.~(\ref{ppmaps:adloss_app}). 
The role of the pre-shock density in hadronic \g-ray images of SNRs is similar to the role of the magnetic field strength in the hard X-ray images. Namely, 
the larger the density the higher the collisional losses of accelerated protons and, therefore, the thinner the radial profiles of \g-ray brightness (Fig.~\ref{fig5}). This effect may be used for estimation of the target-proton density from thickness of \g-ray rims, once the \g-ray observations reach necessary resolution. 


\section{Approximate analytical formula for hadronic $\gamma$-ray images}
\label{pmaps:approx_form}

The approximate formulae (valid close to the shock front) for the radial and azimuthal profiles of the surface brightness of Sedov SNR due to leptonic emission (synchrotron radio, X-rays and inverse-Compton \g-rays) are presented in Papers I and II. The same approach is used here to derive an approximation for the surface brightness profiles of hadronic \g-rays arising from internal structures inside adiabatic SNR. Namely
(Appendix \ref{pmaps:app3}),
\begin{equation}
 S\rs{p}(\varphi,\bar\varrho)\propto
 \varsigma(\Theta\rs{o,eff})\exp\left[-\left( \frac{\epsilon\rs{m}}
 {{\cal E}\rs{max}(\Theta\rs{o,eff})}\right)^{\alpha} \right]
 I\rs{pp}(\Theta\rs{o,eff},\bar\varrho)
 \label{pp_app}
\end{equation}
where  
{\small
\begin{equation}
\begin{array}{ll}
I\rs{pp}
 &\displaystyle \approx 
 {1\over \sqrt{1-\bar \varrho^2}} 
 {1-\bar\varrho^{\sigma(\kappa\rs{pp}+1)}\over \sigma(\kappa\rs{pp}+1)}
 \\ \\  &\displaystyle \times
 \left[1-\frac{\epsilon\rs{m}^{\alpha}\psi\alpha}{{\cal E}\rs{max}^{\alpha}}
 \left(1-\frac{1-\bar\varrho^{\sigma(\kappa\rs{pp}+2)}}{1-\bar\varrho^{\sigma(\kappa\rs{pp}+1)}}
 \frac{\kappa\rs{pp}+1}{\kappa\rs{pp}+2}\right)
 \right],
\end{array}
\end{equation}
}\noindent
$\kappa\rs{pp}={3b}/{2}+(5+s)\kappa\rs{ad}+{1}/{\sigma}-1+\zeta\sigma\rs{pp}(s-1)$, 
$\psi=\kappa\rs{ad}+\zeta\sigma\rs{pp}-3q/2$, $\kappa\rs{ad}\approx 1$, $\sigma$ the shock compression ratio, $\bar\rho=\rho/R$. 
Note, that the azimuthal variation of $I\rs{pp}$ arises only from obliquity dependence ${\cal E}\rs{max}(\Theta\rs{o,eff})$. 

The effective obliquity angle $\Theta\rs{o,eff}$, the azimuth $\varphi$ (measured from the direction of ISMF in the plane of the sky) and the aspect angle $\phi\rs{o}$ are related as
\begin{equation}
 \cos\Theta\rs{o,eff}\left(\varphi,\phi\rs{o}\right)=\cos\varphi\sin\phi\rs{o}.
\end{equation}

Approximation (\ref{pp_app}) is compared with numerical calculations in Appendix \ref{pmaps:app3}. The formula are rather accurate in description of the brightness distribution close to the shock.  
It restores all the properties of the surface brightness profiles of Sedov SNR, revealed in the numerical simulations, 
including dependence on the aspect angle. They are quite useful for qualitative analysis of how different factors affect the 
hadronic image. 
They may also be used as a simple quantitative diagnostic tools for hadronic maps of SNRs. 

For example, the case when the number density is not high, $\zeta\ll 1$. Then, for known $s$, the thickness of the \g-ray rim depends mostly on the compression ratio $\sigma$ and $b$; also on $\alpha$, $\epsilon\rs{m}$ and $q$; the later could be taken zero for an assumption that $E\rs{max}$ is limited by the time of acceleration. 
An opposite case, the product $n\rs{H}(t/1000\un{yrs})\geq 10^5$. Then, the term $\zeta\sigma\rs{pp}$ is dominant in $\kappa\rs{pp}$ and $\psi$ and the thickness of the \g-ray rim may be used to estimate the post-shock number density. 

It is merit to note that our approximate formula reflects also some effects of the non-linear acceleration theory. Namely, in case of efficient proton acceleration, a prominent fraction of the kinetic energy of the shock goes to the relativistic particles. If so, plasma may be described by the adiabatic index $\gamma$ smaller than 5/3. In our formula, the adiabatic index appears through the compression ratio $\sigma$ (Eq.~\ref{sigma-class}) and the parameter $\kappa\rs{ad}$ (Eq.~\ref{calEad-approx}; though $\kappa\rs{ad}$ is close to unity in the range $\gamma=1.1\div 5/3$). Features on images are radially thinner for smaller $\gamma$. Eventual amplification of magnetic field would be accounted through the larger magnetic field compression ratio; however, this ratio does not appear in our formulae because the hadronic images of SNRs do not depend on the strength of magnetic field. 


\section{Discussion and Conclusions}

The spectrum of the TeV \g-rays from SN~1006 may be interpreted as leptonic as hadronic in origin \citep{HESS-SN1006-2010}. The data in the GeV \g-ray range is expected to constrain further this ambiguity. The pattern of surface brightness of SNR also contains important information. In the present paper, we consider possibility that some \g-ray images of SNRs may be hadronic in origin. Namely, we are interested in the images the Sedov SNRs would have if accelerated protons interact with the thermal protons downstream of the shock. The model to synthesize maps of the \g-ray surface brightness of adiabatic SNRs in uniform ISM and uniform ISMF is developed. It includes parameterized description of surface variations of parameters characterizing the injection and acceleration of protons as well as evolution of relativistic protons downstream. The later considers both adiabatic losses of energy and losses due to inelastic collisions. Collisional losses are non-negligible only in case of the large number density of target protons. For example, energy losses of proton with energy 30 TeV due to pp-interactions are comparable to the radiative losses of electrons with the same energy in magnetic field 30 $\un{\mu G}$, is the number density of the target protons is $\sim 10^5\un{cm^3}$. In case the shock moves in the ambient medium with smaller number density, one can consider only the adiabatic energy losses of accelerated protons to model an image and volume-integrated spectrum of SNR. 

The radial thickness of the hadronic \g-ray rim in case of SNR interaction with the large-density ambient material may be used to estimate the density.

The azimuthal variations of the surface brightness of the shell-like adiabatic SNR in hadronic \g-rays is a consequence of the obliquity dependence of the injection efficiency and the maximum energy of accelerated protons. The orientation of the pre-shock magnetic field is the only way the magnetic field influences hadronic images of SNR. If magnetic field is highly turbulent everywhere before the shock of SNR evolving in the uniform ISM then SNR should look as a ring. Nonuniform ISMF (different strength over the SNR surface) does not able to provide any deviation from the ring pattern if information about orientation of ISMF is lost due to turbulence. In contrast, ISM with nonuniform density distribution might provide patterns other than ring, since $K\propto n$. However, it is unlikely that quite symmetrical bilateral structure (as in SN~1006) may appear due to nonuniformity of ISM density because the structure of ISM should be quite special (like a tube, which is not observed around SN~1006). 

Bilateral pattern in hadronic image of the shell-like SNR may therefore be considered as a sign of an ordered ambient magnetic field. If so, the limbs are due to azimuthal variation of either ${\cal K}$ or ${\cal E}\rs{max}$. The most contribution to \g-rays with $\varepsilon\sim 1\un{TeV}$ gives protons with energy $E\rs{p}\sim 8\un{TeV}$. Therefore, if $E\rs{p,max}\gg 8\un{TeV}$\footnote{The maximum energy of accelerated protons is typically expected to be up to the knee in the observed cosmic-ray spectrum at $3000\un{TeV}$. In particular, the hadronic model of TeV \g-rays from SN~1006 suggests $E\rs{p,max}\approx 80 \un{TeV}$ \citep{HESS-SN1006-2010}.} then the bilateral pattern of the hadronic TeV image of SNR reveals the variation of the injection efficiency of protons; in fact, Eq.~(\ref{pp_app}) simplifies to $S\rs{p}(\varphi)\propto \varsigma(\Theta\rs{o,eff}(\varphi,\phi\rs{o}))$. In this case, the hadronic GeV image has to be the same as TeV \g-ray image (differences in GeV and TeV \g-ray images of SNR are signs of the smaller $E\rs{p,max}$ or the contribution from the leptonic emission) and the azimuthal variation of the hadronic \g-ray brightness may be used to derive the obliquity dependence of the proton injection efficiency. Unfortunately, large errors in the present \g-ray data on SN~1006 prevent us from possibility of such an estimate.

\section*{Acknowledgments}

The study was partially supported by 
the program 'Kosmomikrofizyka' of Ukrainian National Academy of Sciences. 
OP acknowledges F.~Bocchino and S.~Orlando for hospitality and many useful discussions.


\appendix
\section[]{Energy losses of protons due to pion production}
\label{pmaps:app0}

In order to obtain the energy losses of a ``single'' proton due to pion production, one needs to integrate the pion source spectra over energies of pions $E_{\pi}$. The pion source spectra is \citep{Schlick-book}
\begin{equation}
 p(E\rs{p},E_{\pi})=cn\rs{H} E_{\pi}\frac{d\sigma\rs{pp}(E_{\pi},E\rs{p})}{dE_{\pi}}H(E\rs{p}-E\rs{th}),
 \label{pploss1}
\end{equation}
where $c$, $n\rs{H}$ are speed of light and the number density of target protons, $d\sigma\rs{pp}(E_{\pi},E\rs{p})/dE_{\pi}$ the differential cross-section for the interaction of two protons, $E\rs{p}$ the energy of incident proton, $H$ denotes the Heaviside step function 
and $E\rs{th}$ the threshold energy of interaction. The energy losses due to inelastic proton collisions is
 \begin{equation}
 -\left(\frac{dE\rs{p}}{dt}\right)\rs{pp}=3\int\limits_{0}^{E_{\pi,max}}dE\rs{\pi}p(E\rs{p},E\rs{\pi}),
 \label{Bloss2}
\end{equation}
where the factor 3 accounts for the production of $\pi^{0}, \pi^{+}$ and $\pi^{-}$ mesons respectively. 
In the $\delta$-function approximation, the differential cross section is given by \citep{Deremer-1986a, Mori-97, Aha-Ato-2000}
\begin{equation}
 \frac{d\sigma\rs{pp}(E_{\pi},E\rs{p})}{dE\rs{\pi}}=\sigma\rs{pp}(E\rs{p})\delta(E_{\pi}-\kappa E\rs{p,kin}),
 \label{sigmapp}
\end{equation}
The proton collision losses are
\begin{equation}
-\left(\frac{dE\rs{p}}{dt}\right)\rs{pp}=3\kappa cn\rs{H}\sigma \rs{pp}(E\rs{p})E\rs{p,kin},
 \label{Bloss3}
\end{equation}
where $\kappa=0.17$ \citep{Aha-Ato-2000} and therefore $3\kappa=0.51$ that restores the coefficient of inelasticity 
$f\approx 0.5$ \citep[Eq.~(3.13) in][]{Aharonian-book}.

\section[]{Evolution of the proton energy spectrum downstream of the shock in Sedov SNR}
\label{pmaps:app1}

We assume that relativistic protons are confined in the fluid elements which advects them from the region of acceleration 
\citep{Reyn-98}\footnote{This means that we do not account for diffusion in the present consideration. This is the reasonable assumption for protons with $E\rs{p}\sim 1\un{TeV}$ (and smaller energies) which give most contribution to emission at TeV (GeV) photons respectively, and 
for the case when the lengthscale of diffusion is proportional to gyroradius. In fact, the gyroradius of the relativistic proton is the same as gyroradius for electron with the same energy; the lengthscale of diffusion for electrons with energy $E\rs{e}\sim 1\un{TeV}$ (and even for higher energies) are similar or  smaller than the lengthscale for advection \citep{ballet2006}.}. 
An individual proton loses energy due to inelastic collisions and adiabatic expansion. 
Downstream distribution of the target protons is proportional to the density distribution given by \citet{Sedov-59} solution, namely it is $\bar{\rho}(r)=\rho(r)/\rho\rs{s}$ where index ``$s$'' marks the immediately post-shock value. The pion production losses are given by (\ref{Bloss3})
\begin{equation}
 -\dot{E}\rs{p,pp}=3\kappa cn\rs{H,s}\bar{\rho}(r)\sigma\rs{pp}(E\rs{p})E\rs{p}
 \label{Bloss31}
\end{equation} 
(here, we neglect difference between total and kinetic energies of proton because we are mostly interested in $E\rs{p}\gg 1\un{GeV}$).
The adiabatic losses are given by \citep{Reyn-98}
\begin{equation}
 -\dot{E}\rs{p,ad}=\frac{1}{3}\frac{E\rs{p}}{V}\frac{dV}{dt}=-\frac{E\rs{p}}{3\bar{\rho}}\frac{d\bar{\rho}}{dt}.
 \label{adloss}
\end{equation}
The equation for losses is
\begin{equation}
 -\frac{dE\rs{p}}{dt}+\frac{E\rs{p}}{3\bar{\rho}}\frac{d\bar{\rho}}{dt}=3\kappa cn\rs{Hs}\bar{\rho}(r)\sigma\rs{pp}(E\rs{p})E\rs{p}.
 \label{tloss}
\end{equation}
In therms of $w=E\rs{p}/\bar{\rho}^{1/3}$ \citep{Reyn-98}, Eq.~(\ref{tloss}) is
\begin{equation}
\frac{dw}{dt}=-c\rs{1}\bar{\rho}w\sigma\rs{pp}(w\bar{\rho}^{1/3}),
\label{tw}
\end{equation}
where $c\rs{1}=3\E{-27}\kappa c n\rs{Hs}\un{cgs}$. 

Let us take the cross section $\sigma\rs{pp}$ in the simple form \citep{Aha-Ato-2000}
\begin{equation}
 \sigma\rs{pp}(E\rs{p})=28.5+1.8\ln{(c_2E\rs{p})} \ \  \un{mb}
 \label{across_app}
\end{equation}
where $c_2=(1\un{GeV})^{-1}$.

The solution of this equation is 
\begin{equation}
 \ln{\left(\frac{c_2E\rs{p}}{\bar{\rho}^{1/3}}\right)}=
 \frac{\ln{\left(c_2E\rs{pi}\right)}+\ln I}{\mu} 
\label{solequation}
\end{equation}
where $E\rs{pi}$ is the initial proton energy produced on the shock at time $t\rs{i}$, 
\begin{equation}
 I(t)=\exp\left[{c\rs{1}\int\limits_{t\rs{i}}^{t}q(t')\mu(t')dt'}\right],
\end{equation}
\begin{equation}
 \mu(t)=\exp\left[{c\rs{1}\int\limits_{t\rs{i}}^{t}p(t'')dt''}\right]
\end{equation}
\begin{equation}
 p=1.8\bar{\rho}, 
\end{equation}
\begin{equation}
 q=-\bar{\rho}\left(28.5+1.8\ln{\bar{\rho}^{1/3}}\right). 
\end{equation}
It follows from (\ref{solequation}) that $E\rs{pi}$ is related to proton energy $E\rs{p}$ at the time of interest $t$ as
$ E\rs{p}={E\rs{pi}}{{\cal E}\rs{ad}^{\mu}{\cal E}\rs{pp}} $
where 
\begin{equation}
 {\cal E}\rs{ad}=\bar{\rho}^{1/3}
\end{equation} 
represents the adiabatic losses and 
\begin{equation}
 {\cal E}\rs{pp}=(c_2E\rs{p})^{1-\mu}I
 \label{ppmaps:Epp}
\end{equation}
represents the energy losses due to inelastic collisions. 

Self-similarity of the task allows us to write $\mu$ and $I$ in terms of the normalized Lagrangian coordinate $\bar a=a/R$
rather than in terms of time $t$. Namely, $I(\bar a)$ and $\mu(\bar a)$ are dimensionless functions and, with the use of \cite{Sedov-59} solutions for uniform ambient medium, $dt'/t=(5/2)x^{3/2}dx$ where $x(t')=R(t')/R(t)$, and therefore 
\begin{equation}
 \mu(\bar{a})=\exp\left[\zeta{\int\limits_{\bar{a}}^{1}x^{3/2}p\left( \frac{\bar{a}}{x}\right) dx}\right],
 \label{mu}
\end{equation}
\begin{equation}
I(\bar{a})=\exp\left[ \zeta\int\limits_{\bar{a}}^{1}x^{3/2}q\left( \frac{\bar{a}}{x}\right)\mu\left( \frac{\bar{a}}{x}\right) dx\right].
\label{i}
\end{equation}
where $\zeta=5tc\rs{1}/2=1.21\E{-6}t_3n\rs{Hs}$, $t_3=t/1000\un{yrs}$. It is clear from here that ${\cal E}\rs{pp}$ is effective only where the number density of target protons is large, at least $n\rs{Hs}\sim 10^6\un{cm^{-3}}$. 

Thus, relations between energies and energy intervals are
\begin{equation}
 E\rs{pi}=\frac{E\rs{p}}{{\cal E}\rs{ad}(\bar{a})^{\mu(\bar{a})}{\cal E}\rs{pp}(E\rs{p},\bar{a})},
 \label{allloss-1}
\end{equation}
\begin{equation}
 \frac{dE\rs{pi}}{dE\rs{p}}=\frac{\mu(\bar a)}{{\cal E}\rs{ad}(\bar{a})^{\mu(\bar{a})}{\cal E}\rs{pp}(E,\bar{a})}.
 \label{allloss-2}
\end{equation}

Let us assume that, at time $t\rs{i}$, a proton distribution has been produced at the shock 
\begin{equation}
N(E\rs{pi},t\rs{i})=K\rs{s}(t\rs{i})E\rs{pi}^{-s}\exp\left[ -\left( \frac{E\rs{pi}}{E\rs{p,max}(t\rs{i})}\right)^\alpha\right],
\label{evolsp}
\end{equation}
where $\alpha$ is constant. 
The conservation equation
\begin{equation}
N(E\rs{p},a,t)=N(E\rs{pi},a,t\rs{i})\frac{a^{2}dadE\rs{pi}}{\sigma r^{2}drdE\rs{p}},
\label{consequation}
\end{equation}
where 
\begin{equation}
 \sigma=\frac{\rho\rs{s}}{\rho\rs{o}}=\frac{\gamma+1}{\gamma-1}
 \label{sigma-class}
\end{equation}
is the shock compression ratio (index ``o'' marks the pre-shock value), and the continuity equation $\rho\rs{o}(a)a^{2}da=\rho(a,t)r^{2}dr$ yield that the energy spectrum evolves downstream as
\begin{equation}
\begin{array}{l}
 \displaystyle
 N\rs{p}(E\rs{p},a,t)=K(a,t)E^{-s}\rs{p}\ \mu(a){\cal E}\rs{pp}(E\rs{p},\bar{a})^{s-1}
 \\ \\ \displaystyle\qquad
 \times\exp\left[ -\left(\frac{E\rs{p}\bar{a}^{3q/2}}{E\rs{p,max}
 {\cal E}\rs{ad}(\bar{a})^{\mu(\bar{a})}{\cal E}\rs{pp}(E\rs{p},\bar{a})} \right)^{\alpha}\right] 
 \end{array} 
 \label{specev}
\end{equation}
with 
$K(a,t)=K\rs{s}(t\rs{i})\bar{\rho}{\cal E}\rs{ad}(\bar{a})^{\mu(\bar{a})(s-1)}$; the evolution of $K$ is self-similar downstream for $K\rs{s}\propto V^{-b}$:
\begin{equation}
 \bar{K}(\bar{a})=K(a,t)/K\rs{s}(t)=\bar{a}^{3b/2}\bar{\rho}(\bar{a})^{1+\mu(\bar{a})(s-1)/3}.
 \label{ksim}
\end{equation} 

We note that if the cross section $\sigma\rs{pp}\approx \mathrm{const}$ then $\mu=1$ and $q=\sigma\rs{pp}\bar \rho$. 

\section[]{Approximations for evolution of some parameters behind the shock}
\label{pmaps:app2}

We interested in approximations of some function $\chi$ downstream and close to the shock, in the form
\begin{equation}
 {\chi}(\bar{a})\approx\bar{a}^{\kappa}.
\end{equation}
By definition
\begin{equation}
 \kappa=\left[ -\frac{a}{\chi(a)}\frac{\partial \chi(a)}{\partial a}\right]\rs{a=R}=\left[- \frac{\partial \ln {\chi(a)}}{\partial  \ln{a}}\right]\rs{a=R}.
 \label{kappa}
\end{equation}
This approach yields for adiabatic losses \citep{xmaps}
\begin{equation}
 {\cal E}\rs{ad}(\bar a)\approx\bar{a}^{\kappa\rs{ad}},
 \qquad 
 \kappa\rs{ad}=\frac{5\gamma+13}{3(\gamma+1)^2},
 \label{calEad-approx}
\end{equation} 
it is valid for $\bar{r}>0.8$ with error less than few per cent. The value of $\kappa\rs{ad}=1$ for $\gamma=5/3$ and is close to unity for $\gamma=1.1\div 5/3$.  

Applying (\ref{kappa}) to Eqs.~(\ref{mu}) and (\ref{i}) one obtains: 
\begin{equation}
 \mu(\bar{a})\approx a^{-1.8\zeta}, \quad I(\bar{a})\approx a^{28.5\zeta}.
 \label{appmu_i}
\end{equation}
With these approximations, Eq.~(\ref{ppmaps:Epp}) are approximately
\begin{equation}
 {\cal E}\rs{pp}(E\rs{p},\bar{a})\approx \bar a^{\zeta\sigma\rs{pp}(E\rs{p})}
 \label{eppapp}
\end{equation}
where the cross section is in units of mb and we used the property $x^{1-\bar a^{-z}}\approx \bar a^{z \ln(x)}$ which is valid for $\bar a\approx 1$. 
Approximation (\ref{eppapp}) is independent of the adiabatic index $\gamma$.

\section[]{Approximate formula for profiles of the hadronic surface brightness of Sedov SNR}
\label{pmaps:app3}

\subsection{Derivation of the formula}

{\bf 1.} 
The \g-ray emissivity due to meson decays is \citep{Aha-Ato-2000} 
\begin{equation}
 q_{\gamma}(\varepsilon)=2\!\!\!\int\limits_{E_{\pi,\min}(\varepsilon)}\!\!\!\frac{q_{\pi}(E_{\pi})dE_{\pi}}{\sqrt{E_{\pi}^{2}-m_{\pi}^{2}c^{4}}} \quad
 \un{\frac{photons}{cm^{3}\ s\  eV}},
 \label{pdspec}
\end{equation}
the pion emissivity 
\begin{equation}
 q_{\pi}(E_{\pi})=\frac{cn\rs{H}}{\kappa}
 \sigma\rs{pp}\left(E\rs{p}'\right)N\left(E\rs{p}'\right).
\end{equation}
On average, energy transferred from protons to pions is $E\rs{\pi}=\kappa E\rs{p,kin}$, 
where $\kappa=0.17$ accounts also for contribution from $\eta$-mesons; 
therefore an average energy of protons mostly responsible for creation of pions with energy $E\rs{\pi}$ is
\begin{equation}
 E\rs{p}'=m\rs{p}c^2+{E_{\pi}}/{\kappa}.
 \label{pmaps:Eppi}
\end{equation}
The minimum energy of pion to create photon is  
\begin{equation}
 E_{\pi,\min}=\varepsilon+m_{\pi}^{2}c^{4}/4\varepsilon. 
 \label{rad:Epimin}
\end{equation} 
Substitution (\ref{rad:Epimin}) into (\ref{pmaps:Eppi}) results in expression for $E\rs{p,min}'$, an average minimum energy of proton to create photon with energy $\varepsilon$. 

\begin{figure}
 \centering
 \includegraphics[width=8truecm]{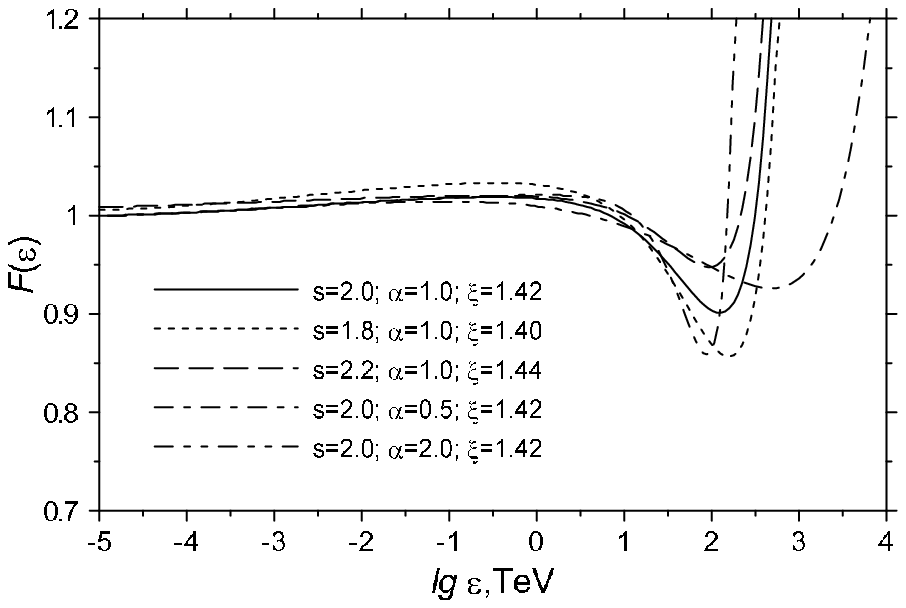}
 \caption{Function $F(\varepsilon)$, Eq.~(\ref{pmaps:Fksi}), for few sets of parameters. $E\rs{p,max}=10^{15}\un{eV}$. 
 Rapid deviation the function from unity at energies above $\varepsilon\sim 0.1E\rs{p,max}$ corresponds to the rapid decrease of \g-ray emissivity.}
 \label{pmaps:fig-ksi}
\end{figure}

Let us introduce an effective energy of protons $E\rs{pm}(\varepsilon)=\xi E\rs{p,min}'(\varepsilon)$ which gives most contribution to \g-rays with energy $\varepsilon$:
\begin{equation}
 E\rs{pm}(\varepsilon)=\frac{\xi}{\kappa}\left(\varepsilon+\frac{m\rs{\pi}^{2}c^4}{4\varepsilon}+\kappa m\rs{p}c^2\right)
 \label{emin}
\end{equation}
and re-write 
Eq.~(\ref{pdspec}) in the form 
\begin{equation} 
 q\rs{\gamma}(\varepsilon)=
 \frac{2cn_{H}}{\kappa}\sigma\rs{pp}(E\rs{pm})N(E\rs{pm})F\rs{p}(\varepsilon,\xi) 
 \label{pmaps:eq5} 
\end{equation}
where 
\begin{equation} 
 F(\varepsilon,\xi) =
 \int \limits_{E\rs{p,min}'(\varepsilon)}^{\infty} 
 \frac{\frac{\sigma\rs{pp}\left(E\rs{p}'\right)}{\sigma\rs{pp}\left(\xi E\rs{p,min}'(\varepsilon)\right)}
 \frac{N\left(E\rs{p}'\right)}{N\left(\xi E\rs{p,min}'(\varepsilon)\right)}\ dE\rs{p}'}{\sqrt{(E\rs{p}'-m\rs{p}c^2)^{2}-m_{\pi}^{2}c^4/\kappa^{2}}}.
\label{pmaps:Fksi}
\end{equation}
$F(\varepsilon,\xi)\approx 1$ is provided by the values $\xi=1.42$, $1.40$, $1.44$ for $s=2$, $1.8$, $2.2$ respectively, in a wide range of $E\rs{p}$ (Fig.~\ref{pmaps:fig-ksi}).  

To the end, Eq.~(\ref{pmaps:eq5}) becomes 
\begin{equation}
 q\rs{\gamma}(\varepsilon)\approx\frac{2cn_{H}}{\kappa}\sigma\rs{pp}(E\rs{pm})N(E\rs{pm})
 \label{pmaps:q} 
\end{equation}
We assume that, close to the shock front, the cross section $\sigma\rs{pp}\approx\mathrm{const}$. The distribution of target thermal protons is proportional to the plasma density, $n\rs{H}\propto  \bar{n}(\bar a)$. Therefore, Eq.~(\ref{pmaps:q}) may approximately be represented with 
\begin{equation}
 q\rs{\gamma}(\varepsilon)\propto \bar{n}(\bar{a})N(E\rs{pm},\bar a).
 \label{pmaps:qa}
\end{equation}

{\bf 2.}
The energy of protons $E\rs{p}$ in a given fluid element at present time 
was $E\rs{pi}$ at the time this element was shocked; they are 
related with (\ref{allloss-1}). 
The behavior of the factors ${\cal E}\rs{ad}$ and ${\cal E}\rs{pp}$ 
close to the shock is given by Eqs.~(\ref{calEad-approx}) and (\ref{eppapp}). 
The distribution $N(E\rs{p})$ is therefore approximately 
\begin{equation}
 \begin{array}{ll}
 N(E\rs{p},\Theta\rs{o})&\propto {\cal K}(\Theta\rs{o})\bar K(\bar a)E\rs{p}^{-s}
 \bar a^{\psi_2} 
 \\ \\ &\times\displaystyle
 \exp\left[ -\left(\frac{E\rs{p} \bar a ^{-\psi}}{E\rs{p,max,\|}{\cal E}\rs{max}(\Theta\rs{o})}\right)^{\alpha}\right] .
 \end{array}
 \label{spa}
\end{equation}
where $\psi_2(E\rs{p})=\zeta\sigma\rs{pp}(s-1)$, 
\begin{equation}
 \psi=\kappa\rs{ad}+\zeta\sigma\rs{pp}-3q/2 
\end{equation}
and we use $\mu=1$ for $\sigma\rs{pp}\approx\mathrm{const}$.

\begin{figure*}
 \centering
 \includegraphics[width=18truecm]{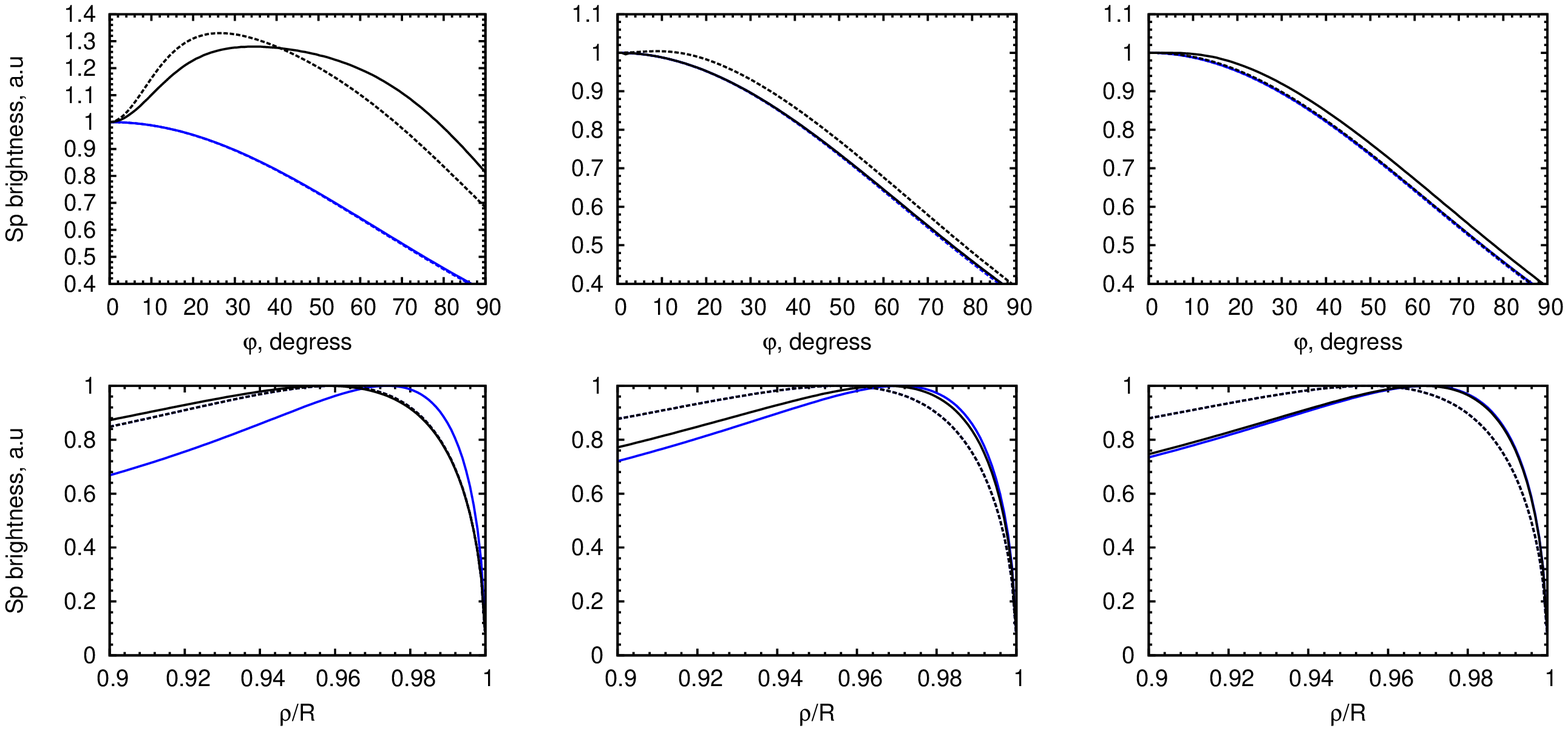}
 \caption{Azimuthal (upper panels) and radial (lower panels) profiles of the hadronic $\gamma$-ray surface brightness $S\rs{p}$ (solid lines) and their approximations (\ref{pazimuth:eq2_fin}) (dashed lines). Calculations are done for $\phi\rs{0}=90^\mathrm{o}$, quasi-parallel injection with $\Theta\rs{K}=\pi/2$, other parameters are the same as in Sect.~\ref{sect-numeric}. Models of $E\rs{max}$: blue line -- ${\cal E}\rs{max}=\mathrm{const}$; black line -- the smooth monotonic variation which provide the variation of ${\cal E}\rs{max}$ in 3.3 times with azimuth (actual representation used is the time-limited model of \citet{Reyn-98} with $\eta=1.5$). The ratio of the photon energy to the maximum proton energy is $\varepsilon/E\rs{p,max}=0.1$ (left), $\varepsilon/E\rs{max}=0.01$ (middle), $\varepsilon/E\rs{max}=0.001$ (right).}
 \label{fig1a}
\end{figure*}

{\bf 3.} 
An `effective' obliquity angle $\Theta\rs{o,eff}$ is related to the azimuthal 
angle $\varphi$ and the aspect angle $\phi\rs{o}$ as (Papers I and II)
\begin{equation}
 \cos\Theta\rs{o,eff}\left(\varphi,\phi\rs{o}\right)=\cos\varphi\sin\phi\rs{o}
\end{equation}
where the azimuth is measured from the direction of ISMF in the plane of the sky.

The surface brightness of SNR is 
\begin{equation}
 S(\bar\varrho,\varphi)=2\int^{1}_{\bar a(\bar\varrho)}q\rs{\gamma}(\bar a) {\bar r \bar r\rs{\bar a} d\bar a\over 
 \sqrt{\bar r^2-\bar \varrho^2}}.
\end{equation}
where $\varrho$ is distance from the centre of the SNR projection, 
$\bar r\rs{\bar a}$ is the derivative of $\bar r(\bar a)$ in respect to $\bar a$.
With the use of Eqs.~(\ref{pmaps:qa}) and (\ref{spa}),
the variation of the meson-decay \g-ray brightness 
is approximately 
\begin{equation}
 S\rs{p}\propto\varsigma(\Theta\rs{o,eff})\exp\left[-\left( \frac{\epsilon\rs{m}(\varepsilon)}
 {{\cal E}\rs{max}(\Theta\rs{o,eff})}\right)^{\alpha}\right]
 I\rs{pp}(\Theta\rs{o,eff},\bar\varrho)
 \label{pazimuth:eq2_fin}
\end{equation}
where $\epsilon\rs{m}=E\rs{pm}/E\rs{p,max,\|}$, 
\begin{equation}
 I\rs{pp}=\int^{1}_{\bar a(\bar\varrho)}\frac{\bar{n}{\bar K} \bar{a}^{\psi_2}\bar r \bar r\rs{\bar a}d\bar a}{\sqrt{\bar r^{2}-\bar \varrho^{2}}}\exp\left[-\frac{\epsilon\rs{m}^{\alpha}}{{\cal E}\rs{max}^{\alpha}} (\bar a^{-\alpha\psi}-1)\right].
 \label{pazimuth:eq2:int}
\end{equation} 

{\bf 4.} 
Let us approximate $I\rs{pp}$. 
Close to the shock front, the density distribution is 
\begin{equation}
 \bar n\approx \bar a^{3\kappa\rs{ad}}
 \label{apprk_n}
\end{equation}
and Eq.~(\ref{ksim}) yields therefore
\begin{equation}
 \bar K(\bar a)\approx 
  \bar a^{\kappa\rs{ad}(2+s)+3b/2}.
 \label{apprk_t}
\end{equation}
In addition \citep{xmaps}: 
$\bar r\rs{\bar a}\approx (1/\sigma)\bar a^{1/\sigma-1}$, $a\approx r^\sigma$,
$\exp\left(-q\rs{*}(a^{-\alpha\psi}-1)\right)\approx 1-q\rs{*}\alpha\psi(1-a)$ and
\begin{equation}
 {\bar r\over\sqrt{\bar r^2-\bar \varrho^2}}\approx {1\over \sqrt{1-\bar \varrho^2}}.
\end{equation}
The integral of interest is therefore 
{\small
\begin{equation}
\begin{array}{l}
I\rs{pp}(\varphi,\bar\varrho)\approx
 \displaystyle
 {1\over \sqrt{1-\bar \varrho^2}} 
 {1-\bar\varrho^{\sigma(\kappa\rs{pp}+1)}\over \sigma(\kappa\rs{pp}+1)}
 \\ \\ \times\displaystyle
 \left[1-\frac{\epsilon\rs{m}^{\alpha}\alpha\psi}{{\cal E}\rs{max}^{\alpha}}
 \left(1-\frac{1-\bar\varrho^{\sigma(\kappa\rs{pp}+2)}}{1-\bar\varrho^{\sigma(\kappa\rs{pp}+1)}}
 \frac{\kappa\rs{pp}+1}{\kappa\rs{pp}+2}\right)
 \right].
\end{array}
\end{equation}
}\noindent
where 
\begin{equation}
 \kappa\rs{pp}=\frac{3b}{2}+(5+s)\kappa\rs{ad}+\frac{1}{\sigma}-1+\zeta\sigma\rs{pp}(s-1). 
\end{equation} 

The formula Eq.~(\ref{pazimuth:eq2_fin}) gives us the possibility to approximate both the azimuthal and the radial brightness profile for $\bar \varrho$ close to unity.

\subsection[]{Accuracy of the approximation}

Fig.~\ref{fig1a} demonstrates accuracy of the approximation (\ref{pazimuth:eq2_fin}) (upper and lower panels for azimuthal, radial profiles). Our calculations shows that this approximation may be used, with errors less than $\sim 20\div 30\%$, in the range of $\bar\varrho$ from $0.9$ to $1$, for azimuth $\varphi$ where $\varepsilon/E\rs{p,max}\leq 0.5,\ 50,\ 0.2$ for $\alpha=1,\ 0.5,\ 2$ respectively (Fig.~\ref{pmaps:fig-ksi}). Accuracy of approximation is higher for smaller $\Theta\rs{K}$ (because the injection term dominates in the azimuthal variation of the brightness) and for smaller variations of ${\cal E}\rs{max}$ (the role of ${\cal E}\rs{max}$ becomes however less prominent for smaller ratio $\varepsilon/E\rs{max}$).

\label{lastpage}
\end{document}